\documentclass[traditabstract]{aa} 
\usepackage{amsmath}
\usepackage{txfonts} 
\usepackage{graphicx}
\usepackage{subfigure}
\usepackage{color}
\usepackage{bm}
\usepackage{natbib}
\usepackage{url}
\usepackage{lscape}
\usepackage{longtable}
\usepackage{rotating}

\newcommand{\simlt}
       {\ifmmode       { \raisebox{-.4em}{$<$}\atop\sim}
          \else        {$\raisebox{-.4em}{$<$}\atop\sim$}
       \fi}

\begin{document} 
\title{Subarcsecond SMA\thanks{The Submillimeter Array is a joint project between the Smithsonian Astrophysical Observatory and the Academia Sinica Institute of Astronomy and Astrophysics and is funded by the Smithsonian Institution and the Academia Sinica.} observations of the prototype Class~0 object VLA1623 at 1.3~mm: A single protostar with a structured outflow cavity ?}
\titlerunning{SMA 1.3-mm observations of VLA1623}
\author{Ana\"elle Maury\inst{1} 
\and Nagayoshi Ohashi\inst{2,3}
\and Philippe Andr\'e\inst{4} 
}
\institute{ESO, Karl Schwarzschild Strasse 2, 85748 Garching bei M\"unchen, Germany
\and Academia Sinica Institute of Astronomy and Astrophysics, P.O. Box 23-141, Taipei 10617, Taiwan
\and Subaru Telescope, National Astronomical Observatory of Japan, Japan
\and Laboratoire AIM, CEA/DSM--CNRS--Universit\'e Paris Diderot, IRFU/SAp, CEA Saclay, 91191 Gif-sur-Yvette, France}
\date{Received 30 September 2011/ Accepted 24 January 2012} 
\abstract 
{
We present 1.3-mm subarcsecond SMA observations of the prototypical Class~0 protostar VLA1623. 
We report the detection of 1.3-mm continuum emission both from the central protostellar component VLA1623 and two additional sources, Knot-A and Knot-B, which have been already detected at longer wavelengths. Knot-A and Knot-B are both located along the western cavity wall opened by the protostellar outflow from VLA1623. Our SMA observations moreover show that these two continuum sources are associated with bright, high-velocity $^{12}$CO(2--1) emission, slightly shifted downstream of the outflow propagation direction with respect to the 1.3-mm continuum emission peaks. 
The alignment of Knot-A and Knot-B along the protostellar outflow cavity, the compactness of their 1.3-mm continuum emission 
and the properties of the associated CO emission suggest that these two sources trace outflow features due to shocks along the cavity wall,
rather than protostellar objects.  
While it was considered as one of the best examples of a close protobinary system so far, the present analysis suggests that the prototypical Class~0, VLA1623, is single on the scales $a>100$~AU probed by our SMA observations. Moreover, we present here the second robust case of compact millimeter continuum emission produced by interactions between the protostellar jet and the envelope of a Class~0 protostar, which suggests a high occurrence of these outflow features during the embedded phase.
} 
\maketitle 

\section{Introduction} 

Understanding the first steps of protostar formation is a major unsolved gap in modern
astrophysics. Observationally, the key to constraining protostar 
formation models lies in high-resolution
studies of the youngest protostars. Class 0 objects, originally 
discovered at (sub)millimeter wavelengths 
are believed to be such very young accreting protostars 
\citep{Andre93, Andre00}. 
Because they are observed only $t < 10^5$~yr after their formation, while 
most of their mass is still in the form of a dense
envelope, Class 0 protostars are likely to 
retain detailed information on the
initial conditions and physics of the collapse phase
(see review by \citealt{Andre00}).  
It is, moreover, believed that the Class~0 phase represents a pivotal stage 
in star formation during which the angular momentum problem \citep{Bodenheimer95} is solved through ejection processes and/or the fragmentation into multiple systems. Indeed, it has been shown that Class~0 protostars drive powerful, highly-collimated protostellar jets \citep{Bontemps96a, Gueth99, Nisini02b}. 
These jets are mostly detected through the swept-up gas forming a molecular outflow, or through a trail of high speed emission knots such as molecular bullets \citep{Bachiller91a}. 
The exact nature of these knots is still unclear because they could result from internal shocks within the jet or be induced by the interaction of the outflowing gas with the surrounding circumstellar medium \citep{Moraghan08}.

VLA~16234-2417 (hereafter VLA1623) was the first Class 0 object to be identified \citep{Andre90a, Andre93}, 
and it is considered to be the prototype of the class. 
The protostar\footnote{In the following, we refer to the system composed of the central protostellar object plus envelope as a protostar.} was originally identified as a strong 1.2-mm continuum source \citep{Andre93} associated with a large-scale CO outflow \citep{Andre90a}, while free-free radio continuum emission, likely to be tracing the shock-ionized base of the jet, was also reported \citep{Leous91}.
High-resolution VLA observations of this object were carried out by \citet{Bontemps97}, revealing 
a string of three candidate HH-like objects (labeled as knots A, B, and C along a western direction) closely aligned 
along the bipolar flow traced by the CO \citep{Andre93} and H$_{\rm{2}}$ emission \citep{Dent95}.
As part of a subarcsecond dust continuum survey of embedded young stellar objects with BIMA at 2.7-mm, 
\citet{Looney00} found two components separated by 1.1\arcsec ($\sim$ 160~AU) at the position 
of VLA1623. 
The eastern 2.7-mm BIMA component is coincident with the centre of the VLA1623 envelope as traced by the larger scale OVRO 2.7-mm map \citep{Bontemps97}.
The 2.7-mm western component is coincident with the first (Knot-A in \citealt{Bontemps97}) of the VLA cm knots. 
While the eastern component, named VLA1623A, is agreed by all to be protostellar in nature, and it most likely traces the central protostellar object of VLA1623, the nature of the secondary component (VLA1623B in the naming system by \citealt{Looney00}, previously  VLA source labeled as Knot-A) remains unclear.
Indeed, \citet{Bontemps97} proposed that it is the first in a series of knots in the jet
from VLA1623, while \citet{Looney00} argued that it is a binary component
of the VLA1623 protostar.
This controversy over whether VLA1623 is a binary protostar has recently been discussed at the light 
of 7-mm VLA observations by \citet{Ward-Thompson11}. 
While they suggest that the millimeter fluxes should mainly be due to dust continuum emission toward both sources, 
they could not firmly 
determine the nature of VLA1623B/Knot-A.
In the following, we adopt the first naming system introduced by \citet{Bontemps97} to refer to the sources in the VLA1623 environment. 

\vspace{-0.2cm}
\section{SMA observations and data reduction} 

We obtained 1.3-mm high angular resolution maps of the prototype Class 0 protostar VLA1623, with the Sub-Millimeter Array (SMA) interferometer \citep{Ho04}. 
Observations were carried out on 2009 July 3 under good weather conditions ($\tau$$\sim$0.1), with the Very Extended configuration (longest baseline $\sim$ 509~m) of the array. 

The central frequency was set to 220.22~GHz in the lower side band (LSB), while the upper side band was centered at 230.3~GHz. High-resolution spectral windows (0.2\,MHz channels) were chosen to cover the $\sim$100\,km\,s$^{-1}$ around the rest frequencies of $^{12}$CO(2--1) at 230.538~GHz, $^{13}$CO(2--1), and C$^{18}$O(2--1). Low-resolution spectral windows (3.25\,MHz channels) were used to cover the rest of the bandwidth. The continuum was built from both the USB and LSB low-resolution spectral windows, excluding a $\sim$50\,km\,s$^{-1}$ spectral window around the e-CH$_{3}$OH(8($-$1, 8)-- 7( 0, 7)) rest frequency (229.8\,GHz), which resulted in an effective line-free bandwidth of 3.2~GHz around a central frequency $\sim$226 GHz.
The baseline solution obtained for Very Extended Configuration on 2009 July 9 was applied, and we performed absolute flux calibration using Callisto observations, while bandpass calibration was obtained by observing the bright quasar 1924-292. Phase and amplitude calibrations during the track were carried out using the quasars 1517-243 (primary calibrator located 15$^{\circ}$ from VLA1623, flux of 1.4~Jy at the time of the observations) and 1625-254 (secondary calibrator located 1$^{\circ}$ from VLA1623, flux of 0.8~Jy at the time of the observations).
Less than 8$\%$ of the data was flagged during the standard reduction procedure, using the software package MIR dedicated to SMA data \citep{Qi11}. 
By comparing the fluxes of the quasars in our calibrated maps with their tabulated fluxes in the SMA database, the final uncertainty in the flux scale was estimated to be less than 15$\%$.

The data were later imaged (inversion and cleaning) using the MAPPING software (part of the GILDAS\footnote{Grenoble Image and Line Data Analysis System, software provided and actively developed by IRAM (\textcolor{blue}{\url{http://www.iram.fr/IRAMFR/GILDAS}})} package). The primary beam FWHM of SMA at 225~GHz is $\sim$50$\arcsec$, and primary beam attenuation has been corrected.
Natural weighting was used for imaging the data, leading to a final full-width at half maximum (FWHM) synthesized
beam-size of 0.7$\arcsec$$\times$ 0.4$\arcsec$, at a position angle of 31$^{\circ}$. 
The 1-sigma rms noise on the final continuum map is 2.1~mJy/beam.
We smoothed the original spectral data by binning 2 consecutive channels, so that the final spectral maps have a spectral resolution 
of 0.5~km\,s$^{-1}$.
The typical 1-sigma rms noise level in the final CO maps is $\sim$ 80 to 120~mJy/beam per channel. 

Here, we only discuss the content and analysis of the continuum and $^{12}$CO(2--1) maps, because interferometric filtering (the shortest baseline is 68~m) and low signal-to-noise ratio in our $^{13}$CO and C$^{18}$O maps prevent any robust analysis without additional shorter spacing data.

\begin{figure*}
\centering
\includegraphics[width=0.98\linewidth,angle=0,trim=0cm 0cm 0cm 0cm,clip=true]{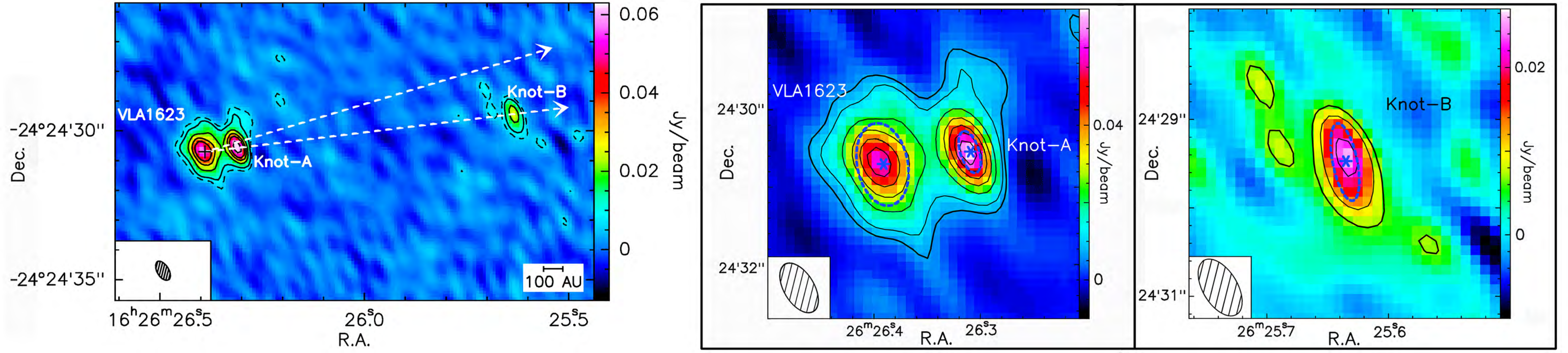}
\caption{{\it{Left panel:}} Wide-field SMA 1.3-mm continuum map towards VLA1623. In this map the beam size is 0.7\arcsec$\times$0.4\arcsec, allowing the three components detected at the positions of VLA1623, Knot-A and Knot-B to be disentangled. The rms noise in the continuum map is 2.1~mJy/beam. Black contours are levels of 6.3~mJy/beam (3$\sigma$-level, dashed contour), then 5$\sigma$ to 20$\sigma$ levels by 5$\sigma$, and ultimately 30$\sigma$ and 40$\sigma$ levels. The two white arrows represent the opening angle ($\sim 30^{\circ}$) of the protostellar outflow according to CO 
low-resolution observations and knots of shocked H$_{\rm{2}}$ emission \citep{Andre90a,Dent95}.
{\it{-- Right Panel:}} Blow-ups of the 1.3-mm continuum map towards the three detected sources. 
The positions (blue stars) and FWHMs (dashed ellipses) of the deconvolved sources modeled (see Table~1) are shown.
}
\label{fig:cont_map_model}
\end{figure*}

\section{Results}
\subsection{SMA 1.3-mm dust continuum map} 

The 1.3-mm continuum map shown in the left hand panel of Figure~\ref{fig:cont_map_model} shows a total 
of three sources detected above the 10-$\sigma$ level. 
The two central sources are coincident with the two components (VLA1623 to the east, Knot-A to the west) previously detected at both 2.7-mm \citep{Looney00} and 7-mm \citep{Ward-Thompson11}.
The third, westernmost source is coincident with the cm source  
Knot-B by \citet{Bontemps97}. 
The separation between the two central continuum sources is 1.2$\arcsec$, i.e $\sim$170~AU assuming a distance 
$D$=139$\pm$6~pc for the $\rho$ Oph cloud\footnote{VLBA \citep{Loinard08} and extinction \citep{Lombardi08} methods for distance determination are in favor of a slightly smaller distance ($\sim$120~pc) for the $\rho$ Oph core.} \citep{Mamajek08}. 
Knot-B is located 10.4$\arcsec$ away of VLA1623, i.e $\sim$1450~AU toward the west.

\begin{figure*}
\centering
\includegraphics[width=0.6\textwidth,angle=0,trim=0cm 0cm 0cm 0cm,clip=true]{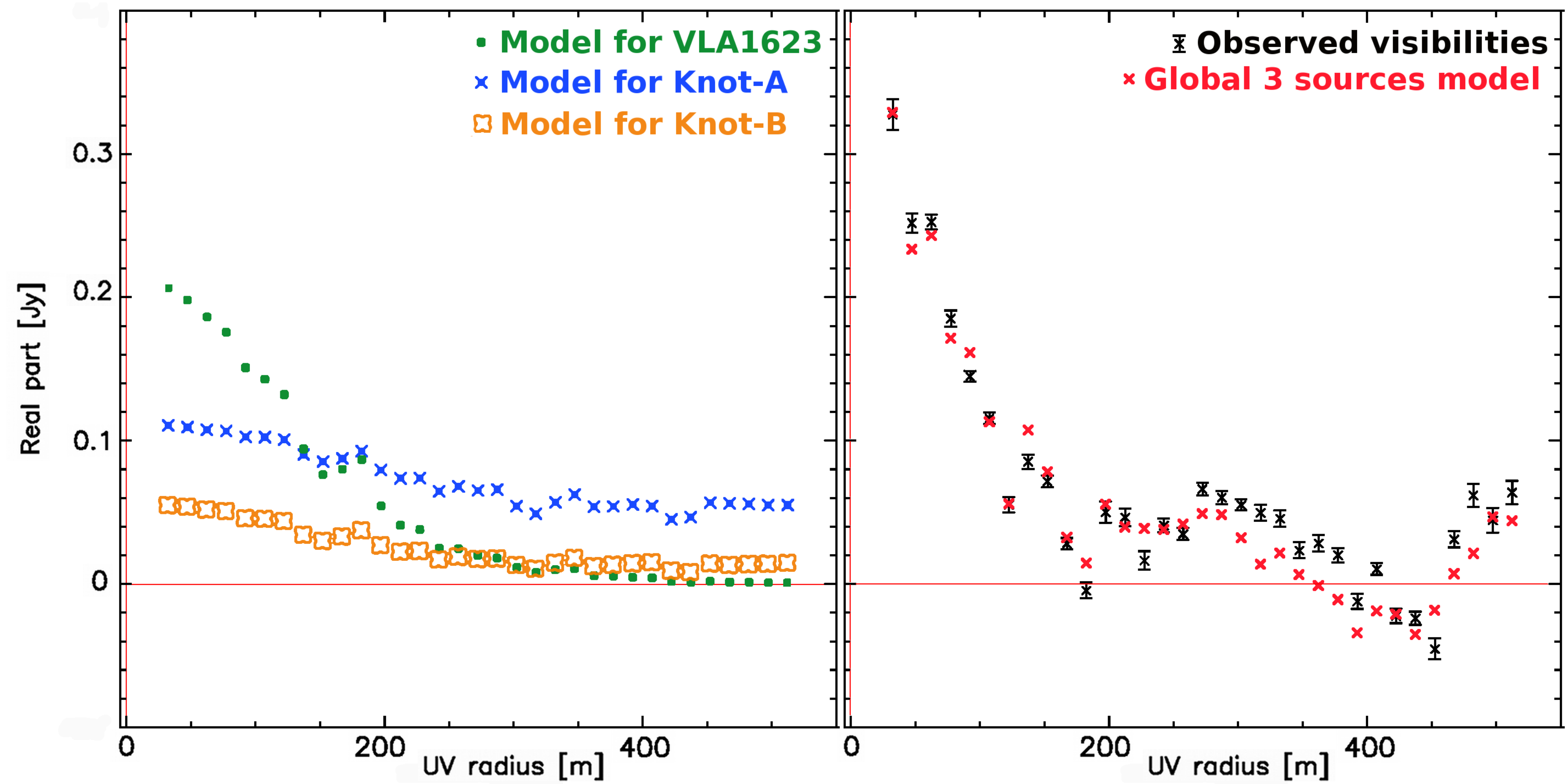}
\caption {The left panel shows, for the three continuum sources, the best-fit model profiles of the visibilities' real parts as a function of projected baseline length. The visibilities for the continuum source modeled at the position of VLA1623 are shown as green dots, the model visibilities towards Knot-A as blue crosses, and the model visibilities reproducing the source at the position of Knot-B are shown as open orange squares. 
The right panel shows the observed profile of the visibilities real parts in our SMA continuum data (plotted as black points with 3$\sigma$ error bars) centered at the position of VLA1623, and the global best-fit model (red crosses) featuring the three gaussian sources shown in the left panel, centered at the same position. The properties of the three continuum sources deduced from this best-fit model are reported in Table~1.
} \label{fig:uvmodel}
\end{figure*}

Table\,1 gives the positions and deconvolved sizes (FWHMs) of the millimeter continuum sources in our SMA map, obtained by fitting the continuum visibilities (in the uv-plane) with a model that includes emission from three 2-D Gaussian sources and shown in Fig.\,\ref{fig:uvmodel}. This modeling work confirms that the profile of VLA1623 is typical of an envelope profile, suggesting that this component dominates the single-dish large-scale continuum emission and is representing the centre of the protostellar envelope of VLA1623.
The continuum profiles of Knot-A and Knot-B, shown in Fig.\,\ref{fig:uvmodel}, show almost no dependency with baseline length, which is typical of much more compact sources. 

The three sources have measured peak and integrated fluxes as reported in Table~1.
The total integrated 1.3-mm emission above the 3$\sigma$ level on the whole primary beam field is 300$\pm$80~mJy. 
We recover 20$\%$ of the total integrated flux of VLA1623 as seen with the IRAM-30m (1500~mJy within a region of diameter 20$\arcsec$, see \citealt{Andre93}). 
Such a low percentage is not surprising considering that our high-resolution observations do not probe the large spatial scales, and therefore filter out most of the extended envelope emission.

\vspace{-0.5cm}
\begin {center}
\begin{table}
\centering \par \caption{Properties of the SMA 1.3-mm continuum sources detected}
\vspace{-0.2cm}
\begin{tabular}{lcccc}
\hline
 {Source}  & VLA1623 & Knot-A & Knot-B \\
 \hline
 $\alpha$ (J2000) $^{(1)}$& {16:26:26.39} & {16:26:26.31} & {16:26:25.63} \\
 $\delta$ (J2000) $^{(1)}$& {-24:24:30.7} & {-24:24:30.5} & {-24:24:29.5} \\
 FWHM $^{(2)}$ & {1.2$\arcsec$$\times$1.1$\arcsec$} & {0.57$\arcsec$$\times$0.24$\arcsec$} & {0.72$\arcsec$$\times$0.27$\arcsec$} \\
 P.A. $^{(2)}$ & 20$^{\circ}$ & 14$^{\circ}$ & 13$^{\circ}$ \\
 F$_{\rm{peak}}$ (mJy/beam)  & {59$\pm$8} & {71$\pm$7} & {26$\pm$4} \\
 F$_{\rm{int}}$ (mJy)  & {175$\pm$25} & {93$\pm$20} & {29$\pm$6} \\
\hline
\end{tabular}
\begin{list}{}{}
\item[$^{(1)}$]{The accuracy on the position of the sources, estimated from absolute positional accuracies of the calibrators observed, is $\sim$0.03$\arcsec$. 
}
\item[$^{(2)}$]{Deconvolved sizes and position angles of the sources (from the modeling of our SMA visibilities, see Fig.\,\ref{fig:uvmodel}.).}
\end{list}
\label{tab:sources}
\end{table}
\end {center} 

\vspace{-2mm}
\subsection{SMA $^{12}$CO(2--1) map}

%
\begin{figure}
\centering
\includegraphics[width=0.99\columnwidth,angle=0,trim=0cm 0cm 0cm 0cm,clip=true]{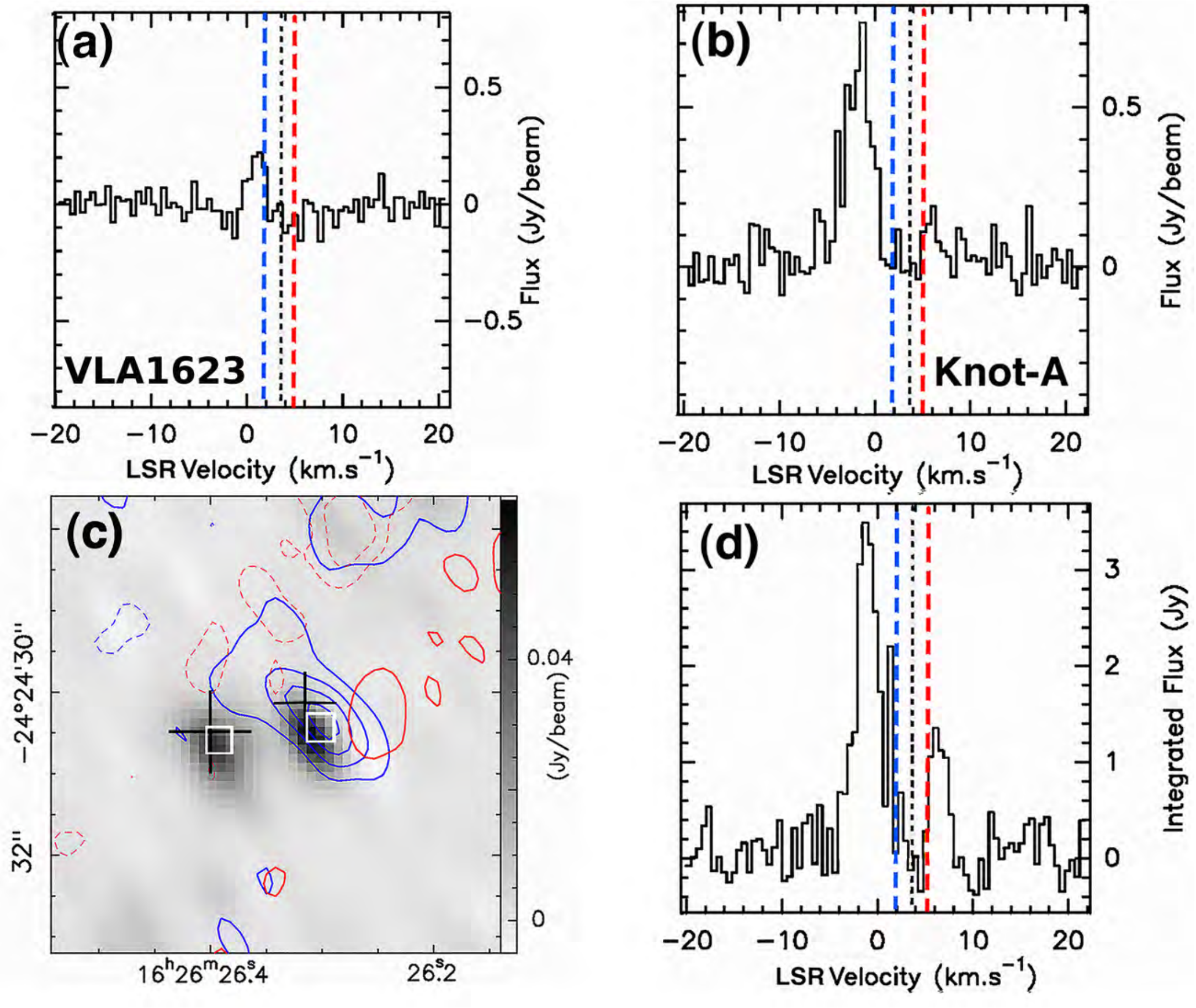}\\
\caption{$^{12}$CO(2--1) emission towards the center of the VLA1623 envelope. Significant CO (above the 10$\sigma$ level) emission, associated with the continuum source Knot-A, is detected.
{\it{(a)}} and {\it{(b)}} show the peak spectrum towards VLA1623 and the peak spectrum of the emission feature associated with Knot-A, respectively. 
In all panels, the central black dotted vertical line indicates the systemic velocity ${\rm{v}}_{\rm{LSR}}\sim$3.8~km\,s$^{-1}$, while the colored dashed vertical lines show the positions of the redshifted and blueshifted components seen in the $^{12}$CO(2--1) single-dish spectrum towards VLA1623 \citep{Andre90a, Dent95}.
{\it{(c)}} shows the 1.3-mm continuum emission map, and contours of $^{12}$CO(2--1) integrated emission: blueshifted from $-$6\,km\,s$^{-1}$ to $+$3.0\,km\,s$^{-1}$, redshifted from $+$4.5\,km\,s$^{-1}$ to $+$10\,km\,s$^{-1}$). The rms noise in the integrated CO emission map is $\sigma \sim$0.15\,Jy/beam\,km\,s$^{-1}$, and the contours shown are levels of 3-$\sigma$ to 11-$\sigma$ by 3-$\sigma$ ($-3\sigma$ dashed contours). The black crosses and the white squares mark the positions of the continuum sources at 2.7-mm \citep{Looney00} and 7-mm \citep{Ward-Thompson11}, respectively.
{\it{(d)}} shows the CO spectrum, integrated over a 3$\arcsec$ box, including the whole extent of continuum emission from both VLA1623 and Knot-A.
}
\label{fig:cont_12co_knotA}
\end{figure}

Since the shortest baseline during our SMA observations was $\sim$68~m (corresponding to a maximum spatial 
scale $\sim$4.7$\arcsec$ probed in the image plane), 
our $^{12}$CO(2--1) map is affected by interferometric spatial filtering.
For example, our SMA map recovers only $\sim$1$\%$ of the high-velocity CO(2--1) emission mapped by \citet{Andre90a} with their single-dish observations.
Thus, not surprisingly, we do not detect the large-scale protostellar outflow observed by \citet{Andre90a}, 
nor the extended emission from the envelope, which are almost entirely resolved out. 
We also suspect that interferometric filtering and/or self-absorption, similar to what is seen in the single-dish observations of VLA1623, 
are responsible for the lack of CO emission in our map near the systemic velocity (${\rm{v}}_{\rm{LSR}}\sim$3.8~km\,s$^{-1}$, \citealt{Narayanan06}). 
Therefore, we stress that the structure seen in our CO map can only trace emission from compact material, at redshifted and blueshifted velocities.

\begin{figure}
\centering
\includegraphics[width=0.95\columnwidth,angle=0,trim=0cm 0cm 0cm 0cm,clip=true]{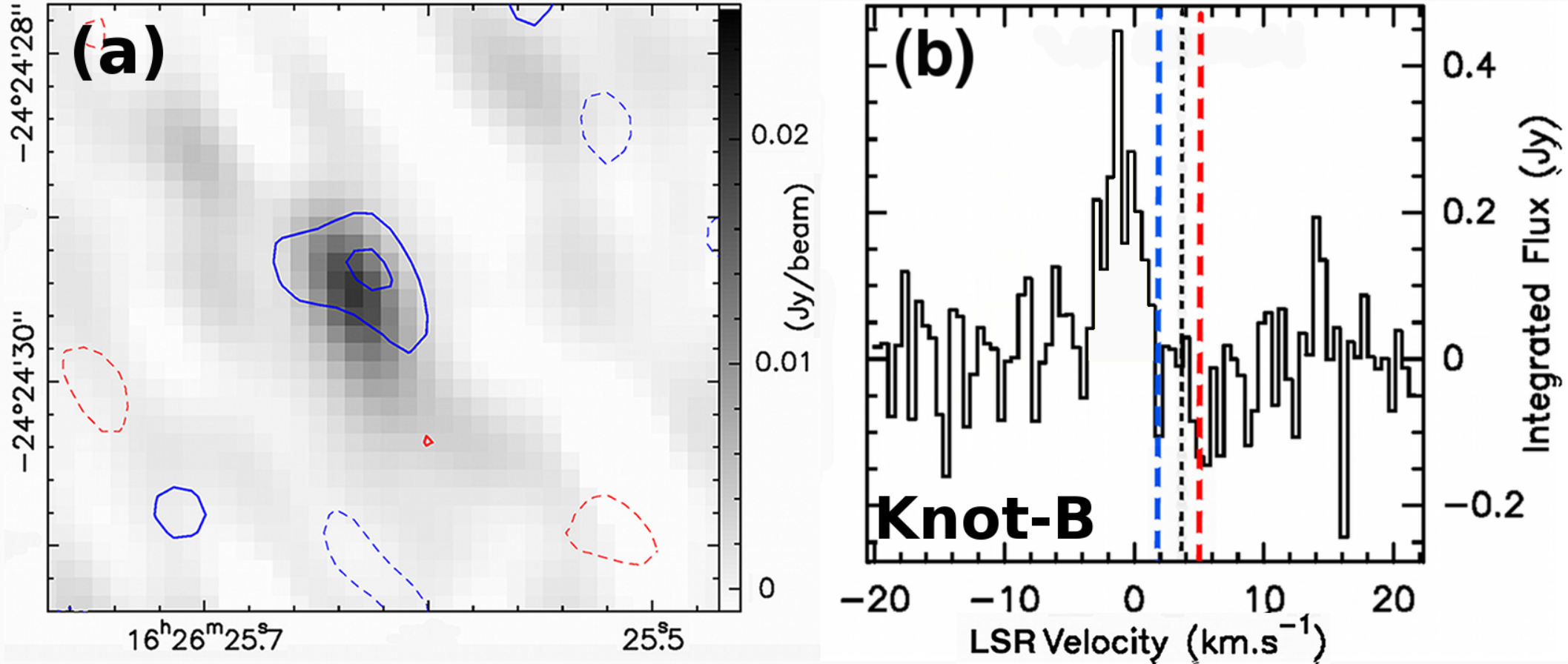}\\
\caption{$^{12}$CO(2--1) emission towards Knot-B. Significant $^{12}$CO (above the $\sim$5$\sigma$ level) emission is detected, as a compact source $\sim$0.2$\arcsec$ away from the 1.3-mm continuum source tracing Knot-B.
{\it{(a)}} shows the 1.3-mm continuum emission map, and the $^{12}$CO(2--1) integrated emission contours (blue from $-$8\,km\,s$^{-1}$ to $+$3.0\,km\,s$^{-1}$, red from $+$4.5\,km\,s$^{-1}$ to $+$10\,km\,s$^{-1}$). The solid contours are levels of 3$\sigma$ and 5$\sigma$, and the dashed contours show the levels of -3$\sigma$, the rms noise being of 0.19\,Jy/beam\,km\,s$^{-1}$. 
{\it{(b)}} shows the integrated $^{12}$CO spectrum of the emission feature associated with Knot-B. The middle black dashed vertical line shows the systemic velocity ${\rm{v}}_{\rm{LSR}}\sim$3.8~km\,s$^{-1}$, while the colored dashed vertical lines show the position of the redshifted and blueshifted components seen in the $^{12}$CO(2--1) single-dish spectrum towards VLA1623 \citep{Andre90a, Dent95}. The noise level is greater towards the spectrum of Knot-B than towards VLA1623 and Knot-A, because this source is located $\sim$10$\arcsec$ away from the phase center chosen for our observations.
}
\label{fig:cont_12co_knotB}
\end{figure}

While tentative ($\sim$3$\sigma$ detection) $^{12}$CO emission is detected towards VLA1623 (see Fig.\ref{fig:cont_12co_knotA}a), 
at slightly blueshifted velocities ($\sim$2\,km\,s$^{-1}$),
the only prominent (detections above the 5$\sigma$ level) emission features in our CO map are 
two small-scale sources, closely associated with Knots A and B. 
Figures~\ref{fig:cont_12co_knotA}c and \ref{fig:cont_12co_knotB}a show the CO integrated emission contours towards Knot-A and Knot-B. 
In both cases, the CO emission is located right beyond the continuum peak, $\sim$0.2--0.4$\arcsec$ downstream along the 
propagation direction of the VLA1623 outflow. 
The integrated spectra shown in Figs.\ref{fig:cont_12co_knotA}d and \ref{fig:cont_12co_knotB}b reveal that, while the CO emission associated with Knot-A has two components (red- and blueshifted) at 
low projected velocities with respect to the systemic velocity of the cloud, only blueshifted CO emission is associated with Knot-B.

Our results therefore suggest that, in a region $\sim$20$\arcsec$$\times$20$\arcsec$ around VLA1623, the brightest compact $^{12}$CO emission structures (shifted from the systemic velocity of the cloud) are associated with Knot-A and Knot-B. 

%

\vspace{-2mm}
\section{Nature of the detected sources}

Our SMA data suggest that, while the 1.3-mm continuum emission from Knot-A and Knot-B is almost point-like, the continuum emission from VLA1623 is partially extended and its visibility curve resembles envelope profiles seen on larger scales (\citealt{Andre93,Ohashi97a}, see electronic Figure\,\ref{fig:uvmodel}). This, and the fact that the continuum emission towards VLA1623 presents the greatest integrated flux, offer further evidence that VLA1623 traces the center of the protostellar envelope seen on larger scales.

The close alignment of both Knot-A and Knot-B with the southwestern outflow cavity wall, as traced by several previous studies 
\citep{Andre90a,Dent95,Yu97}, along with their compactness at millimeter wavelengths and their association with centimeter free-free emission 
first suggests that these two sources are outflow features, i.e shocks between the outflowing material from VLA1623 and the dense ambient material, rather than protostellar components.  
Indeed, it has been suggested \citep{Lee-F01,Cunningham05} that the interaction between circumstellar material and outflowing material 
is likely to create high-density regions in the cavity walls, where the development of a shear layer could also be the site of maser and free-free emission.
Interpretation of Knot-A as a protostellar companion of VLA1623, which is based on the detection of 2.7-mm continuum emission 
towards both sources by \citet{Looney00}, has been proposed when no millimeter continuum emission towards Knot-B had been reported yet.
We stress that, while it would still be statistically 
possible to detect one protostellar companion (such as Knot-A) aligned with the cavity wall of the main protostellar source, the 
detection of two perfectly aligned companions along this axis (such as Knot-A and Knot-B) is highly unlikely. 

Both Knot-A and Knot-B are associated with small-scale, low-velocity CO emission in our SMA map, 
detected at a position shifted by $\sim$0.2$\arcsec$ to $\sim$0.4$\arcsec$ from the continuum emission 
(downward along the outflow propagation direction). 
Such systematic shift between the 1.3-mm continuum and CO emission peaks does not favor a scenario 
where the CO emission would come from a small disk surrounding a young stellar object, because such a disk would also be traced by dust continuum emission at the position of the CO peak.

While the CO single-dish observations of VLA1623 trace only redshifted velocities for the gas towards the western outflow lobe on large scales, the two CO sources associated with Knot-A and Knot-B mostly show blueshifted CO emission. \citet{Murillo10} also report detecting blueshifted CO(3--2) emission along the same redshifted southwestern outflow cavity wall from VLA1623.
Some detections of reverse velocity gas along outflow cavity walls have been reported (see for example the case of Serp-SMM1 by \citealt{vanKempen09b}). These features can be explained as a result of the development of Kelvin-Helmholtz and Rayleigh-Taylor instabilities in outflow cavities (see for example the hydrodynamical simulations by \citealt{Cunningham05}). 
Another explanation would be that the outflow driven by VLA1623 has an axis close to the plane of the sky, as already proposed by \citet{Andre90a}
to explain its low projected velocities, and that it opens a cavity that has a larger opening angle than its angle with the plane of sky.
Such a configuration would lead to overlapping redshifted and blueshifted CO emission detected on both sides of the source, due to the projection of the front and back edges of the cavity lobe along the line of sight \citep{Cabrit86}.

The SMA CO sources that we detect are elongated in the perpendicular direction with respect to the outflow propagation direction. 
This is especially visible in the case of the CO emission associated to Knot-A (FWHM 1.1$\arcsec \times$0.5$\arcsec$, P.A 35$^{\circ}$), because it is 
more extended. Moreover, the CO emission associated to Knot-B has a curved shape, with an apex oriented towards the jet's propagation direction (see Figure~\ref{fig:cont_12co_knotB}b), resembling a bow-shock shape.
This suggests that the $^{12}$CO compact emission detected towards Knots A and B does not trace outflowing material, but is instead due to material interacting with the outflow. Such spatial properties are consistent with a picture where the $^{12}$CO sources originate in small-scale C-type shocks along the walls, as described in the model by \citet{Visser11} and already suggested by \citet{vanKempen10} towards the outflow cavity of HH~46.

Furthermore, the brightness temperature of these CO small-scale sources ($\sim$60~K towards Knot-A, 30~K towards Knot-B, see Figs.\,\ref{fig:cont_12co_knotA} and \ref{fig:cont_12co_knotB}) is higher than the CO brightness temperature from the large-scale outflowing material probed by single-dish observations ($\sim$20~K). 
Such an increase in the CO brightness temperature in outflow cavities is generally interpreted as a signpost of gas heating, due either to outflow shocks \citep{Bachiller99, Hollenbach97} or to ultraviolet photons produced by the protostellar accretion \citep{vanKempen09a}.

Following the new compelling arguments brought by our SMA observations of VLA1623 presented above, we propose that 
Knot-A and Knot-B might represent outflow-features, 
due to shocks between the outflowing gas and the ambient material along the southwestern cavity wall. 
Emission from outflow features in the millimeter continuum maps of young protostars has already been reported by \citet{Gueth03} towards L1157, \citet{Yen10} towards B335, or \citet{vanKempen09a} towards HH~46, on larger scales. 
While further observations of these sources are needed to assess the exact nature of the millimeter continuum emission, it is likely that the emission is due to a blend of free-free emission (from the shock) and dust continuum emission (from the local accumulation of material).
The bright CO emission detected in Knots A and B might also come from column density enhancement and/or 
heating of the gas, similar to what was observed towards IRAS~16293 by \citet{Yeh08}, and towards HH~46 by \citet{vanKempen10} thanks to high-J CO lines emission.
Complementary molecular observations are needed to confirm the exact 
physical processes (shocks, photon heating) responsible for both the CO and continuum emission towards these sources.

These observations provide the clearest example so far of outflow/envelope interactions in the inner few hundred AU of protostars, resulting in compact dust continuum emission. Given the importance of outflows and jets during the early protostellar stages, such features may be common. It therefore demonstrates the need for thorough analysis of the millimeter continuum components detected close to the axes of protostellar jets, since they might be due to outflow features rather than to protostellar companions.

Finally, while it was considered as a good example of a close protobinary system so far, the present analysis suggests that VLA1623 is single on the scales $a>100$~AU probed by our SMA observations. 
This supports the conclusion of \citet{Maury10}, who propose a low rate of small-scale (1000~AU$>a>$100~AU) 
multiplicity for Class~0 protostars, also because VLA1623 is the prototype for protostars at the Class~0 stage.

~\\
{\it{Acknowledgments: The research leading to these results has received funding from the European Community's Seventh Framework Programme (/FP7/2007-2013/) under grant agreement No 229517.}}

\end{document}